\documentclass[
    ,final            
  ]
  {aipproc}

\layoutstyle{6x9}

\def\ghi{$E_{\rm peak}-E_{\gamma}$}
\def\ama{$E_{\rm peak}-E_{\rm iso}$}
\def\yone{$E_{\rm peak}-L_{\rm iso}$}

\def\ep{$E_{\rm peak}$}
\def\epo{$E_{\rm peak,obs}$}
\def\epof{$E_{\rm peak,obs}-F$}
\def\epop{$E_{\rm peak,obs}-P$}
\def\egamma{$E_{\gamma}$}
\def\eiso{$E_{\rm iso}$}
\def\liso{$L_{\rm iso}$}

\begin{document}

\title{Advances on GRB as cosmological tools}


\author{G. Ghirlanda}{
  address={INAF-Osservatorio Astronomico di Brera. Via E. Bianchi 46, I-23807 Merate (LC), Italy}}

\begin{abstract}
  Several interesting correlations among Gamma Ray Bursts (GRB) prompt
  and afterglow properties have been found in the recent years. Some
  of these correlations have been proposed also to standardize GRB
  energetics to use them as standard candles in constraining the
  expansion history of the universe up to $z>6$.  However, given the
  still unexplained nature of most of these correlations, only the
  less scattered correlations can be used for constraining the
  cosmological parameters. The updated \ghi\ correlation is
  presented. Caveats of alternative methods of standardizing GRB
  energetics are discussed.
\end{abstract}

\maketitle


\section{Introduction}

Different cosmological probes seem to point toward a consensus
scenario characterized by a universe which is experiencing an
accelerated expansion. Standard candles (SNIa, e.g. \cite{Astier2006}
), standard rulers (clusters, CMB and BAO, e.g. \cite{Percival2007})
and the angular power spectrum of the CMB (e.g. \cite{Lewis2008}) are
used to probe the Hubble expansion flow up to redshifts 1-2 (by means
of SNIa or clusters) or up to the epoch of recombination (through the
CMB anisotropies).

Gamma Ray Bursts are observed up to very high redshifts: the farthest
is GRB080913C at z=6.7 \cite{Greiner2009}. Among GRBs with known
redshifts, 45\% are at $z>2$ and 8\% at $z>4$. The high GRB luminosity
(10$^{51}$ erg) and their detection in the $\gamma$--ray band makes
them attractive as a potential and complementary cosmological tool to
constrain the cosmological models at $z>2$. However, the problem is
that GRBs are all but standard candles \cite{Bloom2003}: their
isotropic equivalent energetics and luminosities span 3-4 orders of
magnitudes. Similarly to SNIa, it has been proposed to use
correlations between various properties of the prompt emission
\cite{Firmani2006,Firmani2006a} and also of the afterglow emission
\cite{Ghirla2004,Ghirla2004a,Liang2005,Schaefer2007} to standardize
GRB energetics.

\section{Standardizing GRB energetics}

{\bf Isotropic} energies (luminosities) can be computed for GRBs with
measured redshifts and well constrained spectral properties. The
spectrum gives the bolometric fluence $F$ (peak flux $P$) and then
$E_{\rm iso}=4\pi d_{L}(z)^{2} F/(1+z)$ ($L_{\rm iso}=4\pi
d_{L}(z)^{2} P$). These two quantities are strongly correlated with
the rest frame peak energy \ep\ of the $\nu F_{\nu}$ spectrum
\cite{Amati2002,Yone2004}.

We have updated the sample of bursts with known $z$ and spectral
parameters to Jan 2009 (the last being GRB 090102 at z=1.547).  These
are 97 GRBs. The fit of both correlations with a powerlaw gives
\ep$\propto$\eiso$^{0.48\pm0.03}$ and
\ep$\propto$\liso$^{0.4\pm0.03}$. However, due to the large scatter of
the data points $\chi^{2}$ is extremely large (492 and 612 for 95
degrees of freedom for the \ama\ and \yone\ correlations,
respectively). The scatter of the data points is defined by their
distance from the best fit line. By modeling the scatter distribution
with a Gaussian we find a logarithmic dispersion of $\sigma=$0.23 dex
and $\sigma$=0.28 dex for the \ama\ and \yone\ correlation,
respectively.  These scatters are much larger than the statistical
errors associated with the observables $\langle \sigma_{E_{\rm peak}}
\rangle=0.1$, $\langle \sigma_{E_{\rm iso}} \rangle=0.06$, $\langle
\sigma_{L_{\rm iso}} \rangle=0.07$. There is then the possibility that
a third variable is responsible for this large scatter.

\begin{figure}
  \includegraphics[width=1.\textwidth]{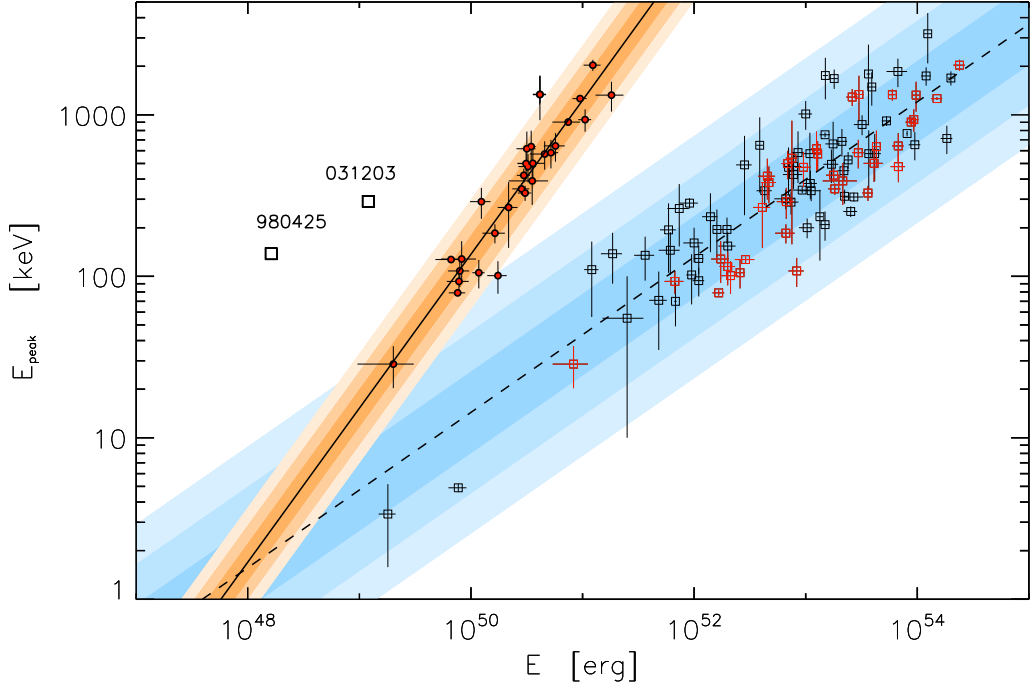}
  \caption{Correlation between the rest frame peak energy and the
    isotropic equivalent energy (open symbols - 97 GRBs) or the
    collimation corrected energy (filled symbols - 29 GRBs). The solid
    (dashed) line is the best fit of the \ghi\ (\ama) correlation. The
    shaded regions show the 1,2,3$\sigma$ dispersion (modeled as a
    Gaussian) of the data points around the best fit lines. Red open
    symbols are the GRBs with a jet break time measurement reported in
    the literature.}
  \label{fg1}
\end{figure}

The large dispersion of the \ama\ and \yone\ correlations prevents
their use to standardize GRB energetics. However, GRBs are thought to
be collimated sources. In the standard GRB model the jetted
outflow should produce a break in the afterglow light curve decay
\cite{Rhoads1997}. The measure of the time of occurrence of this break
$t_{\rm jet}$ allows to infer the jet opening angle $\theta_{\rm jet}$
(under the standard afterglow model assumptions) and to recover the
{\bf collimation corrected} energy $E_{\gamma}=E_{\rm
  iso}(1-\cos\theta_{\rm jet})$.

By collecting available estimates of jet break times from the
literature we found \cite{Ghirla2004} and later confirmed
\cite{Nava2006,Ghirla2006} that the collimation corrected energy is
strongly correlated with \ep. In Fig.\ref{fg1} the \ghi\ correlation
is shown through the 29 bursts having a jet break in their optical
afterglow light curves. We find $E_{\rm
  peak}=E_{\gamma}^{1.04\pm0.08}$ with $\chi^{2}$=37.8 for 27 degrees
of freedom assuming a wind--like profile of the circumburst
medium. This correlation has a dispersion $\sigma=0.07$ which is
consistent with the average statistical uncertainties on \ep\ and
\egamma\ and it is much smaller than the dispersion of the \ama\ and
\yone\ correlations. This result also suggests that the dispersion of
the \ama\ correlation is due to the jet opening angle: by correcting
\eiso\ for $\theta_{\rm jet}$ (for each burst), the scatter of the
\ama\ correlation is reduced.

Due to its tightness the \ghi\ correlation can be used to constrain
the cosmological parameters \cite{Ghirla2004,Ghirla2006}. The \ghi\
correlation, shown in Fig.\ref{fg1}, was found by assuming a
cosmological model (i.e. $\Omega_{M}=0.3$,
$\Omega_{\Lambda}=h=0.7$). It would be a circular argument to use this
particular correlation to constrain the cosmological parameters. We
originally solved this problem by properly accounting for the
dependence of the correlation from the cosmological parameters
\cite{Ghirla2004} or by adopting a Bayesian fitting method
\cite{Firmani2006,Ghirla2006}.

\begin{figure}
  \includegraphics[width=0.8\textwidth,height=0.5\textheight]{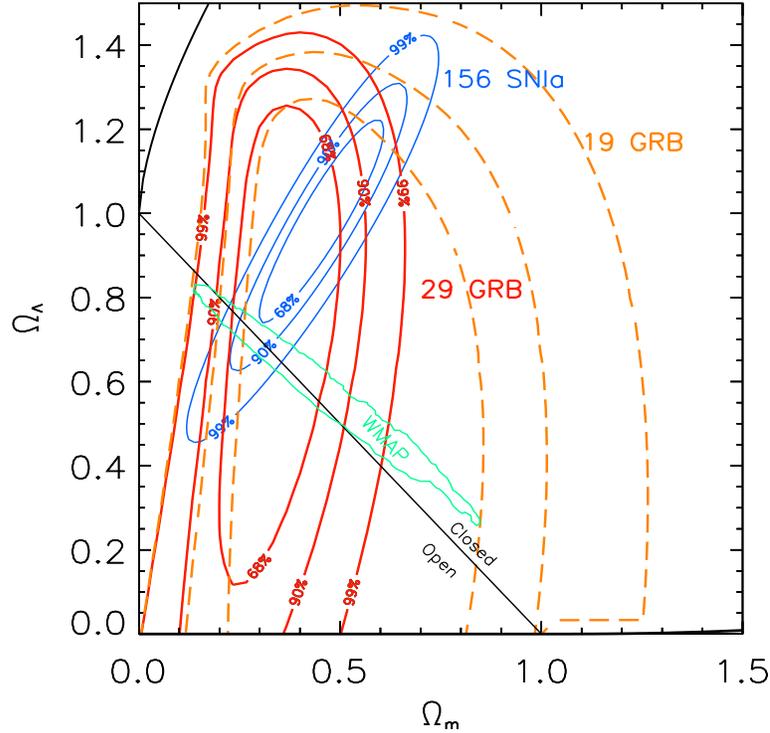}
  \caption{Constraints on the cosmological parameters obtained with
    the \ghi\ correlation updated to Jan 2009 (29 bursts - solid line)
    compared to the previous update (19 GRBs - dashed line -
    \cite{Ghirla2006}).  Also shown are the constraints obtained from
    156 SN type Ia (blue thin line - \cite{Riess2004}) and those from
    the WMAP data (green thin line).}
  \label{fg2}
\end{figure}

Fig.\ref{fg2} shows the cosmological constraints obtained through the
\ghi\ correlation with the most updated sample of 29 GRB. These are
compared with the constraints obtained with the sample of 19 GRBs
\cite{Ghirla2006}. Also in this case we have applied the Bayesian
method that overcomes the circularity problem.

\section{Issues}

The \ghi\ correlation is derived in the standard uniform jet scenario
assuming a constant radiative efficiency and either a uniform
\cite{Ghirla2004} or a wind \cite{Nava2006} circumburst
density profile.  \cite{Liang2005} discovered a completely
empirical correlation \eiso$\propto E_{\rm peak}^{2}t_{jet}^{-1}$
between the three observables which are combined in the \ghi\
correlation. Also the empirical correlation, due to its low scatter,
can be used to derive constraints on the cosmological parameters
\cite{Liang2005}.

The mostly debated issue of the spectral--energy correlations in
general is that they are due to selection effects
\cite{Butler2008}. It has also been claimed that the slope and
normalization of the \ama\ correlation evolve with redshift
\cite{Li2008} Selection effects can be studied in the observational
plane corresponding to these correlations: the \epof\ and the \epop\
planes, where $F$ and $P$ are the bolometic fluence and peak flux.  We
have studied \cite{Ghirla2008,Nava2008} two instrumental selection
effects: the minimum flux required to trigger a burst (trigger
threshold) and the minimum flux to properly analyze its prompt
emission spectrum (spectral threshold). The former has been claimed to
bias the \ama\ correlation \cite{Butler2008}. The sample of 76
bursts (updated to Sept. 2007) with measured redshifts is composed by
GRBs detected by different instruments. For this reason we modeled the
trigger threshold and the spectral analysis threshold of the different
detectors.

We can exclude that the \ama\ correlation is biased by the trigger
threshold. Also we can exclude that the spectral threshold is biasing
the pre-Swift sample.  Instead, the Swift GRB sample (27 events) is
biased by the spectral threshold. This is also due to the limited
spectral energy range of the BAT instrument on-board Swift which
limits the measure of \epo\ in the 15-150 keV energy range.  By
considering sub-samples of bursts at different redshifts, we also
exclude that the \ama\ correlation slope or normalization change.

Recently \cite{Nava2008} added GRBs without redshift in the \epof\ and
\epop\ planes to compare the distribution of bursts with respect to
the two selection effects.  To this purpose a sample of 100 faint
BATSE bursts, representative of a larger population of 1000 objects,
was analyzed. By means of this complete, fluence-limited, GRB sample,
it was found that the fainter BATSE bursts have smaller \epo\ than
those of bright events.  As a consequence, the \epo\ of these bursts
is correlated with the fluence, though with a slope flatter than that
defined by bursts with z.  Selection effects, which are present, are
not responsible for the existence of such a correlation. About six per
cent of these bursts are surely outliers of the \ama\ correlation,
since they are inconsistent with it for any redshift. \epo\ also
correlates with the peak flux, with a slope similar to the \yone\
correlation.  In this case, there is only one sure outlier.  The
scatter of the \epop\ correlation defined by the BATSE bursts of this
sample is significantly smaller than the \epof\ correlation of the
same bursts, while for the bursts with known redshift the \ama\
correlation is tighter than the \yone\ one. Once a very large number
of bursts with \epo\ and redshift will be available, we expect that
the \yone\ correlation will be similar to that currently found,
whereas it is very likely that the \ama\ correlation will become
flatter and with a larger scatter.

One of the main drawback of the \ghi\ correlation for cosmological use
is that it still has few points: the original sample of 15 events
\cite{Ghirla2004} has only doubled since 2004. With the launch of
Swift \cite{Gehrels2004} several jet breaks were expected to be
measured, especially in the X--ray band. Jet breaks should be
achromatic because they are produced by a geometric effect. All the
jet breaks used to compute \egamma\ in the pre-Swift era were,
instead, obtained from the optical light curves. Swift revealed a
complex X--ray afterglow light curve: the early afterglow is often
characterized by a steep decay followed by a shallow phase lasting
thousands of seconds \cite{Burrows2005}. A characteristic break time
is that ending the X--ray shallow phase. \cite{Nava2007} showed that
this time is not a jet break. A possible interpretation is that the
shallow phase is produced by a long lasting central engine activity
\cite{Ghisellini2007} as also supported by the presence of strong
precursors, post-cursors, and X-ray flares in a sizable fraction of
bursts.  Often the X-ray and the optical afterglow light curves do not
track one another, suggesting that they are two different emission
components.

We selected a sample of 33 Gamma Ray Bursts (GRBs) detected by Swift,
with known redshift and optical extinction at the host frame
\cite{Ghisellini2009}.  The de--absorbed and K--corrected X--ray and
optical rest frame light curves are modelled as the sum of two
components: emission from the forward shock due to the interaction of
a fireball with the circum-burst medium and an additional component,
treated in a completely phenomenological way. The latter can be
identified, among other possibilities, as "late prompt" emission
produced by a long lived central engine with mechanisms similar to
those responsible for the production of the "standard" early prompt
radiation.  We find a good agreement with the data, despite of their
complexity and diversity.  Our approach allows us to interpret the
behaviour of the optical and X-ray afterglows in a coherent way, by a
relatively simple scenario. Within this context it is possible to
explain why sometimes no jet break is observed; why, even if a jet
break is observed, it is often chromatic; why the steepening after the
jet break time is often shallower than predicted.

\section{Can we use the \ama\ or \yone\ correlations for Cosmology?}

The use of the \ghi\ correlation for cosmology requires to measure the
redshift, the prompt emission \epo\ and $t_{\rm jet}$. The latter is
the most critical observable: the jet break is typically observed at
1-2 days after the trigger and the light curve needs to be sampled at
much later epochs in order to infer $t_{\rm jet}$ when the afterglow
can be very dim, also because Swift detects higher redshift bursts
than before \cite{Berger2005}.

Therefore, it is interesting to explore if other correlations can be
employed to standardize GRB energetics. Recent attempts
\cite{Amati2008,Kodama2008} tried to use the \ama\ and \yone\
correlations. In this cases, however, one has to take into account
that these correlations are affected by a dispersion which is much
larger than the statistical uncertainty on the data points.

The scatter of these correlations is due to three terms. One is the
statistical uncertainty in the measurements of the parameters
$\sigma_{\rm stat}$.  A second contribution to the scatter comes from
systematic errors $\sigma_{\rm sys}$ which could also have a physical
origin but is difficult to model. A third contribution to the scatter
could finally be due to the cosmological model $\sigma_{\rm cosmo}$:
in the real cosmology its contribution should be minimized. In the
\ama\ and \yone\ correlations the last two terms are dominating the
scatter.

\begin{figure}
  \includegraphics[width=0.8\textwidth,height=0.5\textheight]{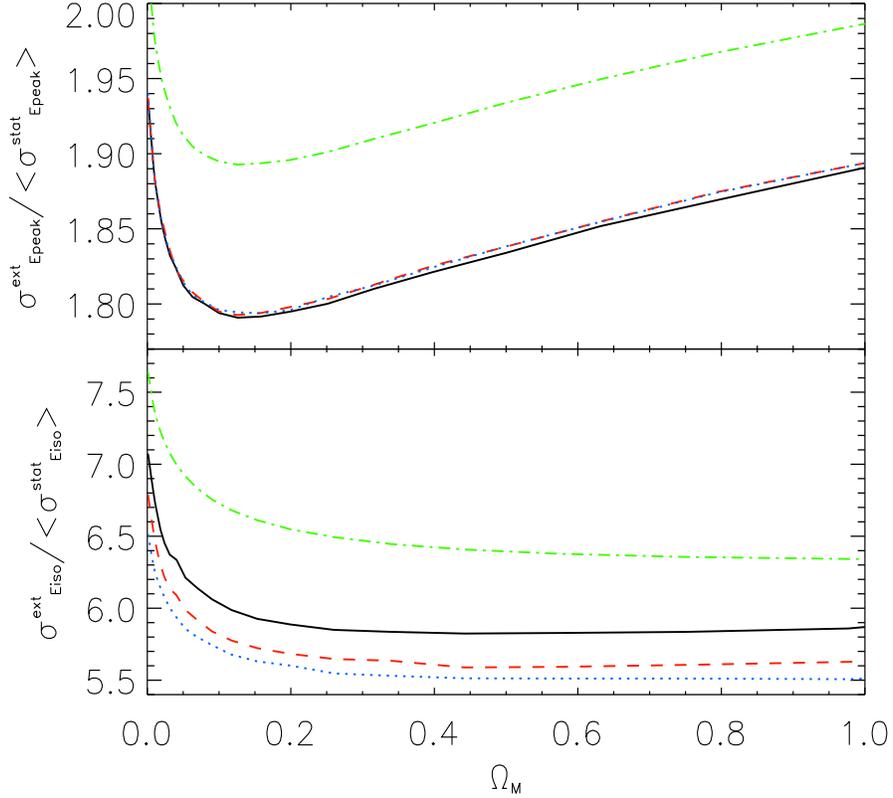}
  \caption{Test of the use of the \ama\ correlation for constraining
    the cosmological parameters. The extra scatter terms of Eq.1,
    i.e. $\sigma_{x,y}$ vs $\Omega_{M}$ are normalized to the average
    value of the statistical error associate to the variable they are
    assigned to. Top panel: the extra-scatter term is assigned to \ep\
    and the results similar to those of \cite{Amati2008} are found. A
    minimum (though weak) is found for $\Omega_{M}\sim 0.15$. The
    solid curve is found with the likelihood function of Eq.1, the
    dotted and dashed curved are found with the symmetric likelihood
    function of \cite{Reichart2001}. The dot-dashed line is found
    with the $\chi^2$ fitting method. Bottom panel: the extra-scatter
    term is assigned to \eiso. This is also the most obvious
    assumption as it is \eiso\ to depend on the cosmological
    parameters. No minimum is found in this case of the extra-scatter
    term for any value of $\Omega_{M}$.}
  \label{fg3}
\end{figure}

\cite{Amati2008} (\cite{Kodama2008}) proposed the use of the \ama\
(\yone) correlation to constrain the cosmological parameters. What is
appealing is the possibility to use a correlation defined by a large
GRB sample (much larger than that defining the \ghi\ correlation) and
to use only on prompt emission observables (\ep\ and \eiso or \liso). 

However, in both correlations the non--statistical scatter 
$\sigma_{\rm  sys}^2+\sigma_{\rm cosmo}^2$ need to be modeled. These terms
(combined) are treated as a free parameter, i.e. the extra--scatter
$\sigma_{\rm ext}$ which is assumed to have a Gaussian distribution
equal for all the data points.

To test the possibility of using the \ama\ correlation for cosmology,
\cite{Amati2008} fit the \ama\ correlation in different
``cosmologies'' (a flat Universe is assumed) and derive, in function
of $\Omega_{M}$, the best fit values of the free parameters, i.e. the
slope $m$ and the normalization $q$ of the correlation and the
extra--scatter term $\sigma_{\rm ext}$. The correlation is fitted with
the likelihood function:

\begin{eqnarray}
\nonumber
\log{P}[m,q,\sigma_{x},\sigma_{y}|(x_{i},y_{i},\sigma_{x,i},\sigma_{y,i})]=\frac{1}{2}\sum_{i} \log[{1\over 2\pi(m^2\sigma_{x,i}^2+\sigma_{y,i}^2+m^2\sigma_{x}^2+\sigma_{y}^2} ]+ & \\
 - \frac{(y_{i}-mx_{i}-q)^2}{m^2\sigma_{x,i}^2+\sigma_{y,i}^2+m^2\sigma_{x}^2+\sigma_{y}^2} & 
\end{eqnarray}

where $x_{i},y_{i},\sigma_{x,i},\sigma_{y,i}$ are the data points
with their statistical errors and $\sigma_{x},\sigma_{y}$ are the
projection of the extra--scatter $\sigma_{\rm ext}$ along the
coordinate axes.

\cite{Amati2008} assume Y=\ep, X=\eiso and set the extra--scatter term
$\sigma_{x}=0$, i.e. they give the (free) extra--scatter only to \ep.
They find that the extra--scatter $\sigma_{y}$ shows a minimum
corresponding $\Omega_{M}\sim 0.1$. Therefore, they apply the standard
procedure to derive constraints on the cosmological models through the
\ama\ correlation.

However, in the \ama\ (\yone) correlations it is \eiso\ (\liso) that
depends on the cosmological parameters (through the luminosity
distance $d_{\rm L}(z|\Omega_{M},\Omega_{\Lambda},H_{0})$). Therefore, the
extra-scatter term should be assigned to X=\eiso,
i.e. $\sigma_{y}=0$. 

For this reason we repeated the same test on the \ama\ and \yone\
correlations and we do not find any minimum of the extra-scatter
$\sigma_{x}$ when it is assigned to \eiso\, which is actually the
variable that depends on the cosmological parameters.  We
verified our results also by (a) adopting the symmetric likelihood
function of \cite{Reichart2001} (having the term $1+m^2$ in the
numerator of the first term of Eq.1); (b) fitting with the least
square method; (c) inverting the order of the fitting variables
(i.e. setting Y=\eiso and X=\ep). Similar results are also found for
the \yone\ correlation. Our results are shown in Fig.\ref{fg3}.

In the \ghi\ correlation (Fig.\ref{fg1}) the scatter of the data
points is already consistent with the statistical errors associated
with \ep\ and \egamma. The only residual scatter is due to the
cosmological model. This is why, without assumptions on the nature
and ``normality'' of the unknown extra--scatter term, the \ghi\
correlation is preferable to standardize the GRB energetics.  The
\ghi\ correlation also proves that most of the scatter of the \ama\
correlation is due to the jet opening angle, which is different from
burst to burst. 

This shows (as also recently demonstrated by \cite{Basilakos2008})
that the \ama\ and \yone\ correlations cannot be used straightforwardly
to constrain the cosmological parameters due to the unknown nature of
the extra--scatter they are affected by.

\begin{theacknowledgments}
  I am grateful to D. Burlon, A. Celotti, C. Firmani, G. Ghisellini,
  M. Nardini, L. Nava, F. Tavecchio for collaborations and
  discussions. ASI is thanked for I/088/06/0 grant. 
\end{theacknowledgments}

\end{document}